\documentclass[reprint,showpacs,showkeys,superscriptaddress,preprintnumbers,amsmath,amssymb,aps]{revtex4-1}
\usepackage[utf8]{inputenc}
\usepackage[T1]{fontenc}

\usepackage{graphicx}
\usepackage{color}
\usepackage{amsfonts}
\usepackage{amsmath}
\usepackage{graphicx}
\usepackage{bm}
\usepackage{amssymb}
\usepackage{times}
\usepackage{dcolumn}
\usepackage{cases}
\usepackage{booktabs}
\usepackage{setspace}
\usepackage{array}
\usepackage{float}
\usepackage{blindtext}
\usepackage{placeins}
\usepackage{physics}

\usepackage{hyperref}
\newcommand{\singlefigwidthlarge}{0.48}

\makeatletter
 \renewcommand\@make@capt@title[2]{%
 \@ifx@empty\float@link{\@firstofone}{\expandafter\href\expandafter{\float@link}}%
 {\textbf{#1}}\@caption@fignum@sep#2\quad}%
 \makeatother
\begin{document}
	
	\title{Topological Contributions to the Anomalous Nernst and Hall Effect in the Chiral Double-Helimagnetic System}
	
\author{Jacob Gayles}

\author{Jonathan Noky}

\author{Claudia Felser}

\author{Yan Sun}
\address{Max Planck Institute for Chemical Physics of Solids, D-01187 Dresden, Germany}

	\date{\today}
	
	\begin{abstract}
	MnP and FeP are show complex magnetic spiral states with geometric phase contributions to the anomalous Hall effect and the topological Hall effect, where both have thermoelectric Nernst counterparts. We use state of the art first principle calculations to show that both Hall and Nernst effects are enhanced in MnP and FeP due to the temperature dependent magnetic texture. At ambient pressure the inversion breaking crystals shows a transition from a double-spiral helimagnetic phase to a topologically trivial magnetic phase. This helimagnetic structure is determined by exchange frustration and the Dzyaloshinskii-Moriya interaction. In the presence of an external magnetic field the magnetic structure transforms into a fan-like phase with a finite spin chirality that gives rise to an effective magnetic field on the order $\sim$1 T. The topological Hall and Nernst effect from this field are comparable or larger to tightly-wound skyrmion systems. Lastly, we show that the topological effects are strongly dependent on the Fermi surface topology, which has strong variations and even sign changes for similar magnetic textures. 
	\end{abstract}
	
	\maketitle

 The anomalous Hall effect (AHE), discovered experimentally over a century ago, was shown in the last few decades to be due to the breaking of time reversal symmetry (TRS) and the splitting of orbital degeneracy \cite{Nagaosa2010}. In collinear ferromagnetic (FM) systems the TRS is broken by the magnetic moment and the orbital degeneracy is lifted by the spin-orbit coupling (SOC) \cite{Xiao2010b}. However, recently there has been evidence for AHE behavior in topological systems that are independent of the net magnetization and the SOC but due to geometric phases \cite{Shindou2001}. For instance, there is the well known Weyl semimetal which produces a finite AHE for two topological band crossings based solely on the distance in momentum space of these crossings (effective magnetic field) \cite{RevModPhys.90.015001,yan2017,Noky2018}. On the other hand, magnetic textures, i.e. skyrmions, in real space can also produce geometrical phases that result in the topological Hall effect (THE) that is independent of the magnetization strength and does not require SOC \cite{Bruno2004,Metalidis2006,Franz2014,Gayles2015}. Analogous to the AHE, the anomalous Nernst effect (ANE) exist in these topological systems under the same conditions when the electric field is exchanged for a temperature gradient which is also coupled with a topological Nernst effect (TNE) in magnetic textures \cite{Zhang2009,Shiomi2013,Kim2018,Fujishiro2018}. 
 
 ANE and AHE-like behavior arises in skyrmion systems due to the topological winding of the magnetic texture, and are termed topological instead of anomalous \cite{Kim2018}. However, these phenomena can exist in systems with finite spin chirality, that do not display skyrmions \cite{Tatara2002,Sharma2018a}. The simplest case is when three antiferromagnetic (AFM) spins in a 120$^{\circ}$ orientation on a triangular lattice that do not lie in the plane will produce an effective magnetic field that is dependent on the solid angle subtended by the three spins \cite{Chen2014,Nayak2016,Yang2017,Guo2017,Narita2017}. In more complex systems such as conical spirals, fan-like, the winding becomes real and deviates from integer values in contrast to skyrmions \cite{Shiomi2012}. Many of the transition metal monophosphides show such exotic spin textures that can be tuned with magnetic field and temperature \cite{Huber1964,Felcher1971,Haggstrgm1982,Dobrzynski1989,Kallel1974}. 
 
 While the AHE is due to the microscopic electronic structure, with the Berry curvature due to the Fermi sea, the ANE is macroscopic property that depends in the statistical average at the Fermi surface \cite{Xiao2006,Xiao2010b}. This leads to a distinction of the AHE and ANE for chiral textures in external magnetic fields, where slight changes in the Fermi surface does not have a large effect on the AHE but can be drastic for the ANE \cite{Sandratskii1986a,Sandratskii1991}. Furthermore, the THE, typically associated with topological magnetic textures of a finite chirality product, is due to momentum scattering at the Fermi surface \cite{Bruno2004}. Here the TNE has a more stringent dependence on the topology of the Fermi surface. Magnetic textures are also dependent on the temperature and not commonly associated with transverse thermoelectric effects but could lead to new scattering dependent phenomena in large temperature gradients due to the variation in phase \cite{Everschor2012,Kong2013,Kim2018}.
 
\begin{figure}[hpt]
		\includegraphics[width=\singlefigwidthlarge\textwidth]{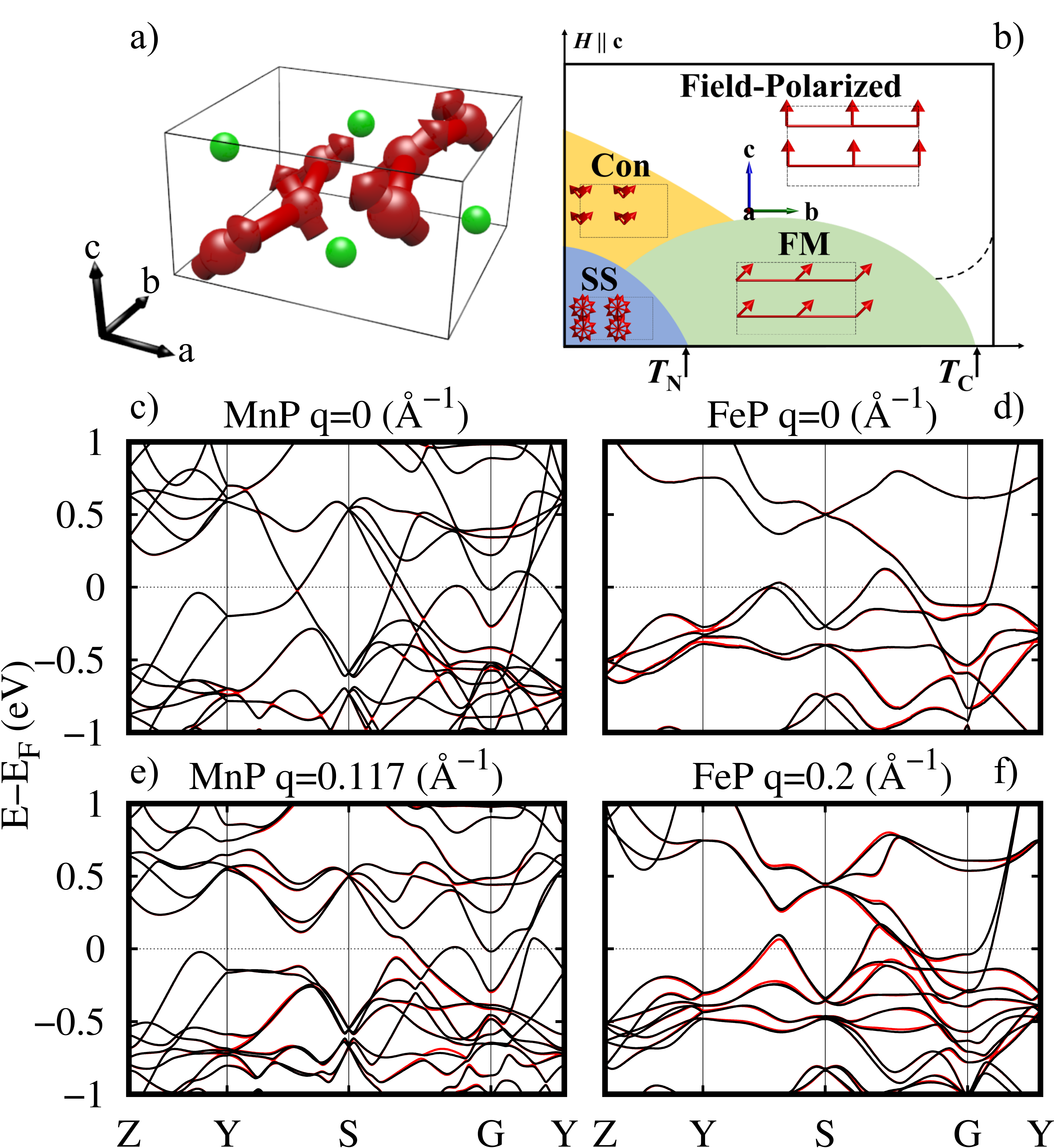}
		\caption{ a) Crystal structure of MnP with the Mn ion in red and the phosphate in green. the bonds show the exchange connection between neighboring Mn. b) A schematic of the magnetic-temperature phase diagram of MnP. Band structure plots with SOC in black and without in red. In c) the collinear FM state of MnP, e) the SS state at $q_a=0.117 \ $\AA$^{-1}$ d) the collinear AFM state of FeP f) the SS sate at $q_a=0.2 \ $\AA$^{-1}$ }
		\label{Phase}
\end{figure}

The phase diagram of the magnetic transition metal monophosphides, with inversion symmetry in the space group {\it Pnma}, show very similar properties \cite{Kitano1964,Obara1980,Dobrzynski1989}. MnP is conventionally known for superconducting properties at low temperatures and high pressures \cite{Xu2017}. In contrast, the ambient pressure MnP at low temperature in zero magnetic field displays a ferromagnetic double helix spin spiral state due to the interactions from ferromagnetic, antiferromagnetic, and antisymmetric exchange \cite{Dobrzynski1989,Sjostrom1992}. In Fig. \ref{Phase}, we show the schematic magnetic field-temperature phase diagram. In the absence of external field, the helimagnetic state transitions to a ferromagnetic state at $T_\mathrm{N}$= 47 K and displays paramagnetic behavior above $T_\mathrm{C}$=291 K \cite{Sjostrom1992}. Whereas, the magnetic crystal FeP displays an antiferromagnetic double-helix ground state, half the period length of MnP, with weak ferromagnetism that only has a transition from the helimagnetic state to the paramagnetic state at $T_\mathrm{N}$=120 K \cite{Sjostrom1992}. However, in both systems there is a transition to a conical state in the presence of an external magnetic field when applied perpendicularly to the spin spiral rotation (see Fig. \ref{Phase}). At larger magnetic fields this state becomes field polarized. These magnetic structures deviate from the adiabatic regime and strictly influence the electronic structure, which in turn should distinguish the intrinsic AHE and ANE and related geometrical responses.

In this work we focus on the distinctive features in the electronic structure that depends intimately on the magnetic structure, which we calculate using first-principle methods. The exotic magnetic structures are calculated self-consistently taking advantage of the generalized Bloch theorem where SOC is applied as a perturbation to calculate the intrinsic AHE and ANE. We calculate finite contributions to both effects in the absence of SOC that is significantly smaller than with SOC. Furthermore, the AHE shows comparable values for different magnetic structures, where in contrast the ANE is strongly dependent on the magnetic structure in magnitude and sign. Conversely, the topical phenomena arise in the these systems within a non-integer winding due to the cone angle subtended by the spiral magnetic states. These phenomena are calculated in both ferromagnetic spirals of MnP and the antiferromagnetic spirals of FeP and are comparable to that of skyrmion systems. The antiferromagnetic FeP displays these effects due to the weak ferromagnetism and the canting of the spins. This work offers valuable insight into the contributions to the Hall and Nernst effect in non-trivial magnetic textures. 

{\it Methods} We employ density functional theory (DFT) calculations of bulk MnP and FeP using the full potential linearized augmented plane wave method as implemented in the J\"{u}lich DFT code \texttt{FLEUR}~\cite{fleur}, and the Perdew-Burke-Ernzerhof parametrization of the exchange-correlation potential \cite{Perdew1996}. From the DFT electronic structure we construct 64 Wannier functions from the {\it p}-states of the phosphorous and {\it d}-states of the transition metal ions. The constructed Wannier \cite{Mostofi2008} function for each magnetic spin spiral takes a Hamiltonian of the form $H=H_0+H_{\mathrm{M}}(\bf{q})$ . The magnetic part of the Hamiltonian is $H_{\mathrm{M}}(\bf{q})$, where the unit direction of the spin at site $i$ is determined by ${\mathbf n}_\mu=(\cos({\bf q\cdot R}_{i,n} + {\mathbf \varphi}_i)\sin\theta,\sin({\bf q\cdot R}_{i,n} + {\mathbf \varphi}_i)\sin\theta, \cos\theta)$. The cone angle $\theta$ is the angle between the spins and the rotation axis, and ${\mathbf \varphi}_i$ is the added phase at a given site. The spiral vector {\bf q} determines the length and direction of the spiral with the rotation angle of a spin at ${\bf R}_{i,n}$ defined as $\phi_{i,n}={\bf q\cdot R}_{i,n}$. 

Due to the {\bf q}-dependent electronic structure that preserves the generalized Bloch theorem, $\mathcal{T}_nH= H\mathcal{T}_n$, the addition of SOC would break the Bloch symmetry \cite{Herring1966a,Sandratskii1991}. $\mathcal{T}_n$ is the generalized translation that combines a spin ration and a lattice translation. Therefore we add SOC as a perturbation, $\Delta \varepsilon_{\mathbf{k},n}(\mathbf{q})=\bra{\psi_{\mathbf{k},n}(\mathbf{q},\mathbf{r})}H_{SO}\ket{\psi_{\mathbf{k},n}(\mathbf{q},\mathbf{r})}$, for the unperturbed spin spiral state $\psi_{\mathbf{k},n}(\mathbf{q},\mathbf{r})$ \cite{Heide2009a}. This pertubation is sufficient in that the exchange fields, $\sim$1 eV, are significantly larger than the spin orbit interaction, where in MnP and FeP the spin-orbit field is 50 and 59.8 meV, respectively. With the SOC perturbation we are able to calculate all relevant spin-orbit contributions. 

In the double-helix monophosphides the spiral propagates along the a-axis with the spins rotating in the b-c plane. The monophosphides form in an orthorhombic crystal structure that breaks inversion symmetry and TRS due to the magnetic moments. The experimental lattice parameters taken of MnP (FeP) are a=5.92 (5.80), b=5.26 (5.21), and c= 3.17 (3.10) \AA~\cite{Forsyth1966}. In MnP the spins are canted due to the Dzyaloshinskii-Moriya interaction (DMI) along the a-axis in the bc-plane showing a phase, ${\mathbf \varphi}_i$, difference between two spins in the unit cell. However, the DMI in antiferromagnetic FeP leads to weak ferromagnetism producing a canting between two antiferromagnetic spins, with the size of the DMI 2.12 and -1.30 meV in MnP and FeP, respectively. The DMI is known to arise in systems that break both inversion symmetry and TRS and is due to the relativistic SOC.

Due to the orthorhombic crystal structure the magnetic crystalline anisotropy is large and prefers the c-axis as the easy axis. Our calculations show the anisotropy energy to be $\Delta_{\mathrm{MCA}}$= 1 and 0.6 meV between the ac and bc directions in MnP. In FeP the anisotropy is weaker and prefers spins to point along the b axis with $\Delta_{\mathrm{MCA}}$= 0.2 meV. The DMI, $E_{\mathrm{DM}}=\sum_i\mathbf{D}_i(\hat{\mathbf{m}})\cdot(\hat{\mathbf{m}}\times\partial_i\hat{\mathbf{m}})$, is also highly anisotropic preferring to point along the c-direction causing a canting angle (phase shift $ \tau_i$) of 2$^{\circ}$-6$^{\circ}$ between neighboring spins. Whereas in FeP the DMI causes a weak ferromagnetism between the antiferromagnetic spins with a phase difference of $\sim$169$^{\circ}$. The alternating bonds along the c-axis cause the DMI vector to reverse sign where the phosphorus ion acts as the broken inversion centre \cite{JSjostrom1992}. 

 	\begin{figure*}[hpt]
		\includegraphics[width=\textwidth]{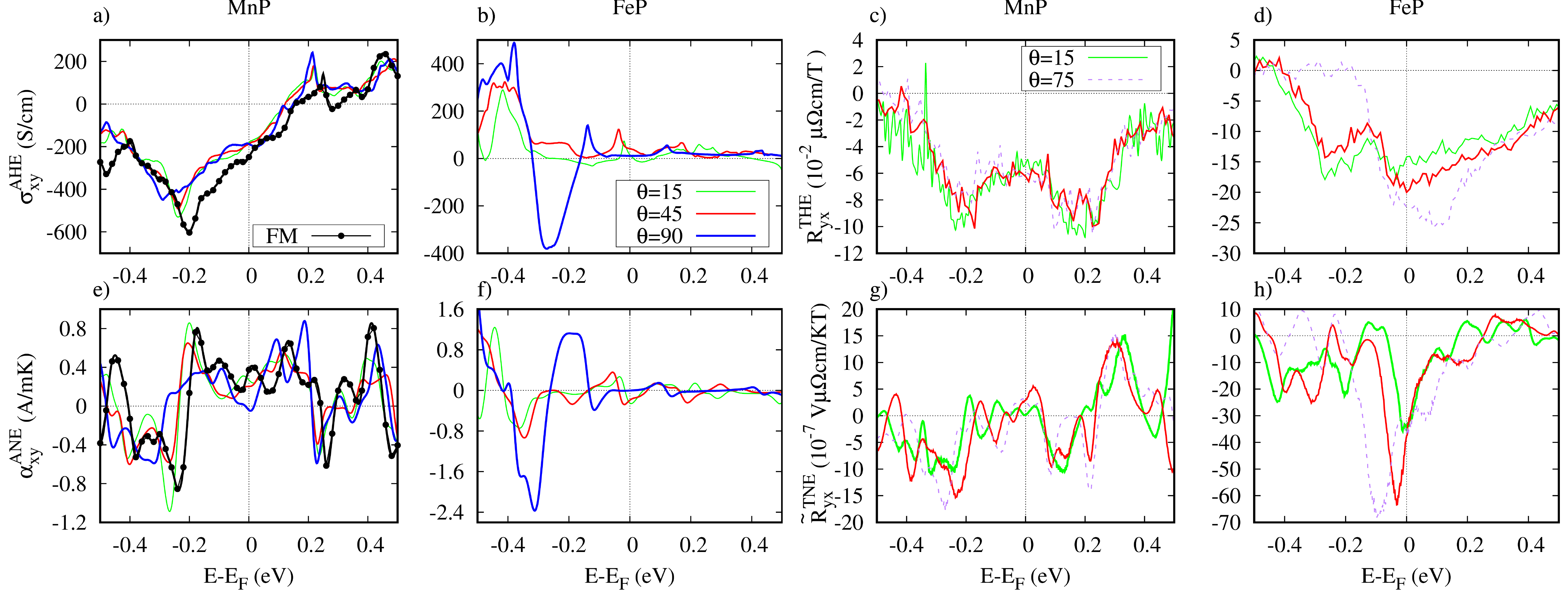}
		\caption{ The top panels show the Hall responses and the bottom show Nernst responses. In the top panels a) and b) show the AHE in MnP and FeP, respectively for different cone angles. c) and d) show the THE for MnP and FeP, respectively. In the bottom panels e) and f) show the ANE for different cone angles in MnP and FeP calculated at T=100K. In g) and h) the topological Nernst effect is shown for MnP and FeP respectively. }
		\label{anoms}
	\end{figure*}
In Fig. \ref{Phase} c) and d) we plot the band structure of MnP with SOC in black and without in red for the collinear and spin-spiral (SS) states, and correspondingly for FeP in Fig. \ref{Phase} e) and f). The effect of the spiral on the system is to break the energy degeneracy at two k-points, $\epsilon(\mathbf{k})\neq \epsilon(-\mathbf{k})$. In the collinear ferromagnetic MnP there are many crossings close the Fermi energy where a gap opens with the addition of SOC ( Fig.\ref{Phase} c)). Furthermore in e), we show the effect of experimental finite $q_a=0.117 \ $\AA$^{-1}$. Here most of the crossing points of the ferromagnetic system are gapped and the result of SOC is to gap the valence bands from the conduction band forming a semimetal. In the case of the collinear AFM FeP, the band structure is shown, which is not seen experimentally but good for comparison to the SS state. Specifically there is a Dirac point lying 0.5 eV above the Fermi energy that is split by the spiralization and weak ferromagnetism. These subtle changes in the electronic structure have significant consequences for the transport phenomena. \
From the Wannier interpolated Hamiltonian we are able to calculate all Berry curvature, $\Omega$, related effects. Firstly the Kubo formula is used to calculate the momentum-space Berry curvature \cite{Nagaosa2010}:\begin{equation}\label{1}
 \Omega_{ij}^{n}=\mathrm{Im}\sum_{m\neq n}\frac{\bra{n}\frac{\partial H}{\partial k_i}\ket{m}\bra{n}\frac{\partial H}{\partial k_j}\ket{m}-(i\leftrightarrow j)}{(\varepsilon_n-\varepsilon_m)^2}.
\end{equation}
$\Omega_{ij}^{n}$ is the $ij$-th component of the momentum Berry curvature of the $n$-th band, where $\ket{n}$ and $\varepsilon_n$ is the eigenstate and eigenvalue of $H$. The AHE is calculated as the k-space integration, $\sigma_{ij}^{\mathrm{AHE}}=-\frac{e^2}{\hbar}\sum_n^{occ}\int\frac{d^3k}{(2\pi)^3}\Omega_{ij}^{n}$ \cite{Nagaosa2010}. Contrary to the AHE the ANE arises in systems when the external field is replaced by a temperature gradient. The work of Xiao et al. shows that the ANE can also be calculated from the momentum space Berry curvature \cite{Xiao2006}:
\begin{equation}\label{2}
\begin{split}
\alpha_{ij}^{\mathrm{ANE}} = \frac{e}{T\hbar}\sum_n & \int\frac{d^3k}{(2\pi)^3}\Omega_{ij}^{n}[(\varepsilon_n-E_F)f_n \\
& +k_BT\mathrm{ln}(1+\mathrm{exp}\frac{\varepsilon_n-E_F}{k_BT})].
\end{split}
\end{equation}
In the above equation, $E_F$ is the Fermi energy and $T$ is the temperature and $f_n$ is the Fermi-Dirac distribution. Furthermore, both effects should follow Onagers reciprocal relations, specifically an applied electric field should result in a transverse temperature gradient. 

We compared the full self consistent collinear ferromagnet with SOC to that of the "spin-spiral" state with $q_a$=0 and $\theta$=0 and adding SOC as a perturbation. The perturbative scheme reproduces the self consistent calculations for an energy range above -0.5 eV, with a slight shift in the energy position of the peaks below. This serves a check of the validity of the perturbation and the electronic structure that the perturbative scheme reproduces the AHE and ANE of the fully self-consistent calculations. Contrary to the AHE, the topological responses show to be independent of the SOC and do not require this pertubation.

A real space magnetic texture, e.g. a skyrmion, also produces a transverse voltage in an applied electric field, that is also coupled with a thermoelectric counterpart. These phenomena differ from the AHE and ANE, in that orbital degeneracy breaking by the SOC is replaced by the real space variations of the magnetic texture. For slowly varying skyrmionic textures with weak SOC the magnetization distribution of the skyrmion is, $\Omega_{\mathbf{RR},ij}=\frac{1}{2} \hat{\mathbf{m}}\cdot(\partial_{\mathbf{R}_i}\hat{\mathbf{m}} \times \partial_{\mathbf{R}_j}\hat{\mathbf{m}} )$. This continuous magnetization texture acts as an effective magnetic field in real space with a sign change of the Lorentz force on spins of opposite sign \cite{Bruno2004}. This effect is usually termed the topological Hall effect due to the topology of the skyrmion, however in this the monophosphides there are no skyrmions but this fan-like structure which produces a finite effective field. 

Analogous to the formulation of the intrinsic Hall and Nernst effect, we formulate for the THE and TNE in magnetic textures. In this, we neglect the effect of the weak SOC in these systems and assume an adiabatic evolution of the electron quasiparticle of a single band within the Boltzmann transport theory using a constant relaxation time approximation \cite{Franz2014,Gayles2015}. Contrary to the intrinsic effects this transport depends only on the Fermi surface topology. The diagonal conductivity is determined with a constant relaxation time, $\tau$, is given by:
 \begin{equation}\label{3}
\sigma_{xx}=\frac{e^2}{VN}\sum_{\mathbf{k}n}\tau\delta(E_F-\varepsilon_{\mathbf{k}n})(v_{\mathbf{k}n}^{x})^2,
\end{equation}
with $V$the volume of the unit cell and $N$ the number of $\mathbf{k}$ points in the Brillouin zone, and $v_{\mathbf{k}n}^{i}$ is the group velocity in the $i$ direction. We take $\tau= \alpha \eta^{-1}$, where $\eta$ is the density of states and $\alpha$ is a constant. Analogous to the AHE, the band dependent momentum space distribution of the spin dependent ordinary Hall effect in an effective field, $B_e$, can be expressed as:
%\begin{equation}\label{4}
%\begin{split}
% \tilde\Omega_{ij}^{n}= & \pm B_{e}\tau^2\delta(E_F-\varepsilon_{\mathbf{k}n}) \\
% & \times[(v_{\mathbf{k}n}^{x})^2m^{yy}_{\mathbf{k}n}-(v_{\mathbf{k}n}^{x})(v_{\mathbf{k}n}^{y})m^{xy}_{\mathbf{k}n}].
%\end{split}
%\end{equation}
\begin{equation}\label{4}
\begin{split}
\sigma_{ij}^{THE}(\pm B_e)= & \pm B_{e}\frac{-e^3}{VN}\sum_{\mathbf{k}n}\tau^2\delta(E_F-\varepsilon_{\mathbf{k}n}) \\
 & \times[(v_{\mathbf{k}n}^{x})^2m^{yy}_{\mathbf{k}n}-(v_{\mathbf{k}n}^{x})(v_{\mathbf{k}n}^{y})m^{xy}_{\mathbf{k}n}].
\end{split}
\end{equation}
With the sign change determined by the spin. In equation~\ref{4} $m^{ij}_{\mathbf{k}n}=\partial^2\varepsilon_{\mathbf{k}n}/(\hbar^2\partial k_i\partial k_j)$ is the inverse effective mass tensor. Consequently the TNE can be calculated using the Mott relation:
\begin{equation}\label{5}
\begin{split}
\alpha_{ij}^{\mathrm{TNE}}(\pm B_e) = -\frac{1}{e}\int d\varepsilon\frac{\varepsilon-\mu}{T}\sigma_{ij}^{THE}(\varepsilon)\frac{\partial f}{\partial \mu}
\end{split}
\end{equation}
The total topological Nernst constant is due to both spins and calculated as:
\begin{equation}\label{6}
\begin{split}
\tilde R_{yx}^{\mathrm{TNE}}(B_e)=\frac{\alpha_{xy}^{\mathrm{TNE},\uparrow}(B_e)-\alpha_{xy}^{\mathrm{TNE},\downarrow}(B_e)}{\sigma_{xx}^2B_e}.
\end{split}
\end{equation}
 The topological Hall resistivity, $\rho_{yx}^{THE}(B_e)$, also follows in a similar manner, where $\alpha^{\mathrm{TNE}}$ is replaced by $\sigma^{\mathrm{THE}}$. Within this approximation $\sigma_{xy}^{THE}=\sigma^2R_{yx}^{THE}B_e$ and $\alpha_{xy}^{TNE}=\sigma^2\tilde R_{yx}^{TNE}B_e$ depend only on the effective field and the constant, where the constant is parameter free and can be calculated purely from the electronic structure. The effective field can be calculated numerically on a discrete lattice, $B_e=\frac{e}{h}\sum_i\hat{\mathbf{m}}(\mathbf{r})\cdot[\hat{\mathbf{m}}(\mathbf{r}+\delta_i)\times \hat{\mathbf{m}}(\mathbf{r}+\delta_{i+1})]$. Here $\delta_i$ is the vector connecting nearest-neighbor sites, where the magnetization is summed over the magnetic unit cell. 
 
 	\begin{figure}[hpt!]
		\includegraphics[width=\singlefigwidthlarge\textwidth]{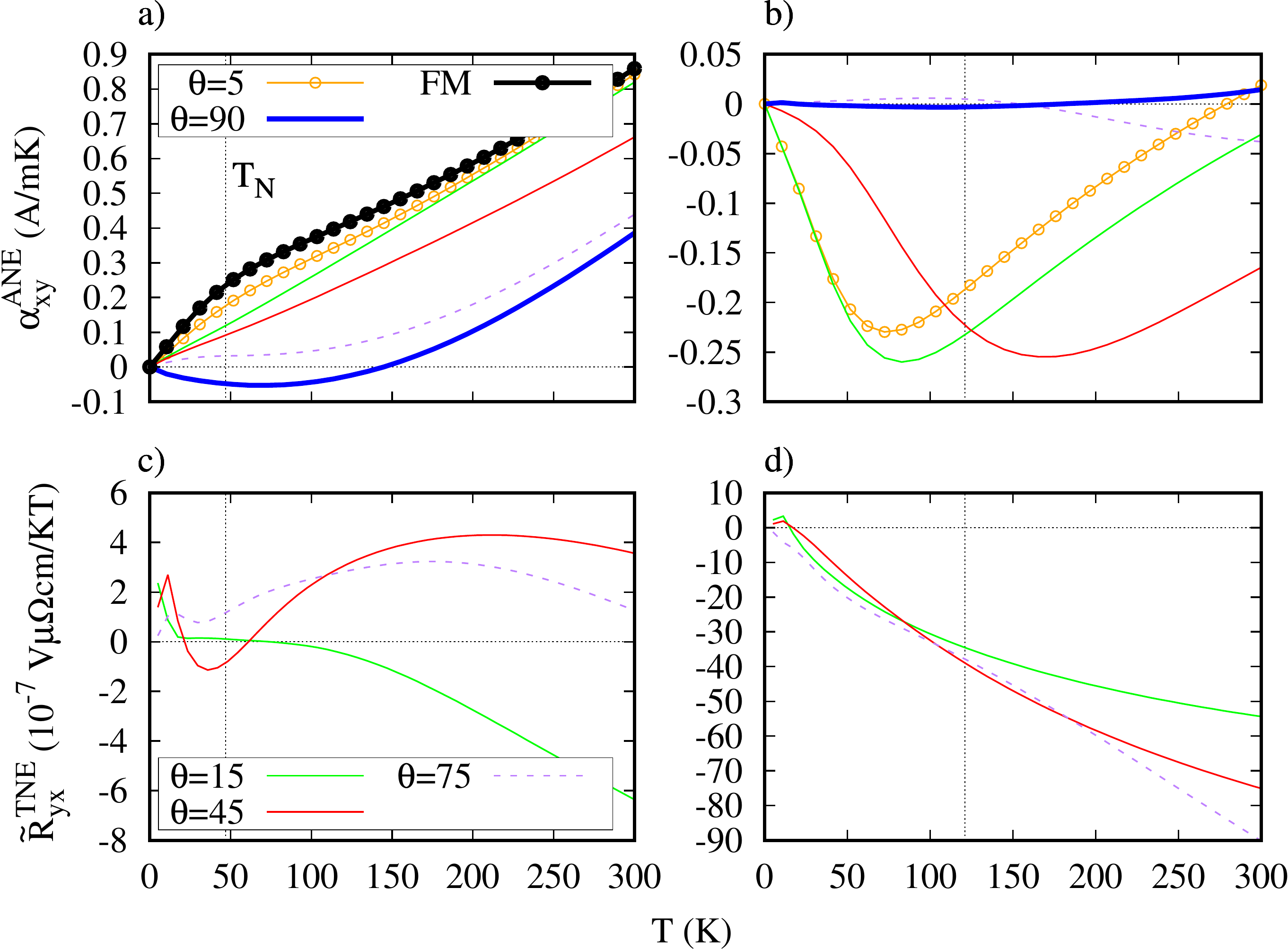}
		\caption{ In the top panels a) and b) show the temperature dependence of the ANE for different cone angles in MnP and FeP respectively. The bottom panels show the TNE for cone angles with finite effective field in MnP c) and FeP d). The dashed line shows the transition temperature at zero magnetic field}
		\label{tnernst}
	\end{figure}

 {\it Results} The result of Shiomi {\it et al.} measure an AHC of 50 S/cm at resistivities, $\rho<10^{-5}~\Omega {\mathrm {cm}}$ \cite{Shiomi2012}. We believe this reduction is due to scattering mechanisms which is beyond the scope of this paper to understand the intrinsic Berry phase related effects \cite{Weischenberg2013}. At smaller resistivities their results show an increase in the AHE to $\sim$300 S/cm which is comparable to our results. In Fig. \ref{anoms} a) and b) the anomalous Hall conductivity is shown close to the Fermi energy in MnP and FeP respectively, for different cone angles at the experimental spiral length. As expected, in MnP the AHE does not show a considerable difference in the spiral phase compared to that of the ferromagnetic state. Whereas in antiferromagnetic FeP the helical spiral, $\theta=90^{\circ}$ (blue curve), there is a sharp change in the AHE as the cone angle is decreased. The helical spiral in FeP is very close to the antiferromagnetic state with nearly an equal population of majority and minority states at the Fermi energy. Furthermore, we calculated the AHE without spin-orbit coupling in the different phases, where the spiralization can lead to a spin-orbit like term breaking the orbital degeneracy of energy levels. Here in both systems the $\theta=90^{\circ}$ shows a zero contribution for the entire energy range as expected. For $0<\theta<90^{\circ}$ there is finite but small AHE that does not reach values larger than $\approx$40 S/cm around the Fermi level. In addition, there is a finite $\sigma_{ij}^{AHE}$ for all transverse components, however we focus on the $xy$ that is perpendicular to the effective magnetic field and an order of magnitude large then the other components.

 Contrary to the AHE in MnP, the ANE has a stringent dependence on the magnetic texture. In Fig. \ref{anoms} e) and f) we show the ANE in MnP and FeP with values reaching a maximum of 0.8 and -2.4 A/mK around the Fermi energy. The figures show the ANE calculated at T=100 K. In the MnP the ANE changes sign at the Fermi energy for the case of $\theta=90^{\circ}$ as compared to the ferromagnetic case and the other conical spirals. In addition, the effect is sizable and should be measurable in experiments. At low temperatures the thermoelectric properties can be related by the Mott formula, $\sim\frac{\partial \sigma}{\partial \varepsilon}$ \cite{Xiao2006}. This shows the explanation to the sign change of the ANE, where although the AHE does not change in size for the different conical spirals, the slope does change for the case of $\theta=90^{\circ}$. Furthermore in the antiferromagnetic helical spiral FeP the slope of the AHE is constant and leads to zero ANE effect. Whereas, for the conical spirals the AHE peaks just below the Fermi level, which constitutes a large ANE at the Fermi level.
 
 The topological responses in this adiabatic regime depend purely on the topology of the Fermi surface. In Fig.~ \ref{anoms} c) and d) we show the graphs of the topological Hall constant in MnP and FeP for magnetic structure with finite effective fields, i.e. $\theta$=15$^{\circ}$ (green), 45$^{\circ}$ (red), and 75$^{\circ}$ (dashed purple). In both systems the THE is comparable to that of skyrmionic systems, where the effective field is also comparable to that of skyrmionic systems ranging between 0.5-1 T. In MnP the topological Hall constant does vary strongly with different magnetic structures. However, in FeP the curves show similar trends for different cone angles with a distinguishable difference as a function of the Fermi energy. This is due to the canting of the antiferromagnetic helix into a ferromagnetic state. 
 
In Fig. \ref{anoms} g) and h) the TNE is plotted as a function of the Fermi energy in MnP and FeP at $T$=100 K. In Fig. \ref{anoms} h) one sees a smooth variation of the peak at the Fermi energy with the size of the conical spiral. The values in MnP show to be about an order of magnitude larger that of skyrmions in MnGe, $\tilde R_{yx}^{TNE}\approx$ 6.9 10$^{-8}$V$\mu\Omega$cm/KT, and two orders of magnitude larger in FeP \cite{Shiomi2013,Fujishiro2018}. This is due to smaller conductivities in the monophosphides which increase the topological Nernst constant. However the effective field in MnGe is an order of magnitude larger than in MnP and this would lead to comparable responses. 

In Fig.~\ref{tnernst} we show the temperature dependent ANE (top panels) and the TNE (bottom panels) in MnP and FeP. In MnP the temperature dependent Nernst effects stongly depend on the cone angle of the spiral. Below the transition temperature, $T_{\mathrm{N}}$, the ANE is positive and increasing in the ferromagnetic state and all cone angles except for $\theta$=90$^{\circ}$ (blue) which is negative and increases well above $T_{\mathrm{N}}$. Above $T_{\mathrm{N}}$ the ferromagnetic state should be resolved in zero external field, however the presence of an external magnetic field extends the $T_{\mathrm{N}}$ for the conical spirals (see Fig.~\ref{Phase} b)). The TNE also shows to vary in sign as a function of the cone angle and temperature. In FeP the ANE shows to be zero for in the helical antiferromagnetic spiral and finite for cone angles less than $\theta=90^{\circ}$. Nearly all curves show a minimum below $T_{\mathrm{N}}$ and increase at higher temperatures. Conversely, the TNE in FeP shows to increase monotonically in magnitude with the increase in temperature. 

In conclusion we have explored the Hall effects and Nernst effects in double helix spirals with finite cone angle of MnP and FeP. We see that the magnetic structure of these monophosphides determines the Fermi surface topology which has drastic effects on the anomalous Nernst and topological Nernst effects. While both effects are sizable and experimentally measurable the topological Nernst effect is substantially large in comparison to skyrmion systems. We believe this work will further the understanding of the topological responses in magnetic fields.
	
We thank Yuriy Mokrousov, Frank Freimuth and Jakub \v{Z}elezn\'{y} for fruitful discussions. This work was financially supported by the ERC Advanced Grant No. 291472 ``Idea Heusler``and ERC Advanced Grant No. 742068 ``TOPMAT``. We also gratefully acknowledge Max Planck Computing and Data Facility (Garching, Germany) for providing computational resources.

\end{document}